\documentclass[twocolumn,superscriptaddress,showpacs,prl]{revtex4}
\usepackage{graphicx}
\usepackage{dcolumn}
\usepackage{natbib}

\newcommand{\kpnn}{K^0_L\rightarrow\pi^0\nu\bar{\nu}}
\newcommand{\kpp}{K^0_L\rightarrow\pi^0\pi^0}
\newcommand{\kppp}{K^0_L\rightarrow 3\pi^0}
\newcommand{\kcppp}{K^0_L\rightarrow\pi^+\pi^-\pi^0}
\newcommand{\kpln}{K^0_L\rightarrow\pi l \nu (l=e,\mu)}

\newcommand{\kgg}{K^0_L \rightarrow \gamma \gamma}
\newcommand*{\PUSAN}{%
Department of Physics, Pusan National University, Busan, 609-735 Republic of Korea}
\newcommand*{\SAGA}{%
Department of Physics, Saga University, Saga, 840-8502 Japan}
\newcommand*{\DUBNA}{%
Laboratory of Nuclear Problems, Joint Institute for Nuclear Research, 
Dubna, Moscow Region, 141980 Russia}
\newcommand*{\SOKENDAI}{%
Department of Particle and Nuclear Research, 
The Graduate University for Advanced Science (SOKENDAI), Tsukuba, Ibaraki, 305-0801 Japan}
\newcommand*{\TAIWAN}{%
Department of Physics, National Taiwan University, Taipei, Taiwan 10617 Republic of China}
\newcommand*{\KEK}{%
Institute of Particle and Nuclear Studies, 
High Energy Accelerator Research Organization (KEK), Tsukuba, Ibaraki, 305-0801 Japan}
\newcommand*{\OSAKA}{%
Department of Physics, Osaka University, Toyonaka, Osaka, 560-0043 Japan }
\newcommand*{\YAMAGATA}{%
Department of Physics, Yamagata University, Yamagata, 990-8560 Japan}
\newcommand*{\CHICAGO}{%
Enrico Fermi Institute, University of Chicago, Chicago, Illinois 60637, USA }
\newcommand*{\NDA}{%
Department of Applied Physics, National Defense Academy, Yokosuka, Kanagawa, 239-8686 Japan}
\newcommand*{\RCNP}{%
Research Center of Nuclear Physics, Osaka University, Ibaraki, Osaka, 567-0047 Japan}
\newcommand*{\KYOTO}{%
Department of Physics, Kyoto University, Kyoto, 606-8502 Japan}
\newcommand*{\IHEP}{%
Institute for High Energy Physics, Protvino, Moscow region, 142281 Russia}
\newcommand*{\GOMEL}{%
Scarina Gomel' State University, Gomel', BY-246699, Belarus}
\newcommand*{\ARIZONA}{%
Department of Physics and Astronomy, Arizona State University, Tempe, Arizona, USA}
\newcommand*{\RIKEN}{%
RIKEN SPring-8 Center, Sayo, Hyogo, 679-5148 Japan}

\begin{document}
\date{April 24 2008}

\title{
Search for the Decay $K^0_L \rightarrow \pi^0 \nu \bar{\nu}$}

\author{J.~K.~Ahn}\affiliation{\PUSAN} 
\author{Y.~Akune}\affiliation{\SAGA} 
\author{V.~Baranov}\affiliation{\DUBNA}
\author{K.~F.~Chen}\affiliation{\TAIWAN}
\author{J.~Comfort}\affiliation{\ARIZONA} 
\author{M.~Doroshenko}\altaffiliation{Present address: \DUBNA}\affiliation{\SOKENDAI} 
\author{Y.~Fujioka}\affiliation{\SAGA} 
\author{Y.B.~Hsiung}\affiliation{\TAIWAN} 
\author{T.~Inagaki}\affiliation{\SOKENDAI}\affiliation{\KEK} 
\author{S.~Ishibashi}\affiliation{\SAGA}
\author{N.~Ishihara}\affiliation{\KEK}
\author{H.~Ishii}\affiliation{\OSAKA} 
\author{E.~Iwai}\affiliation{\OSAKA}
\author{T.~Iwata}\affiliation{\YAMAGATA} 
\author{I.~Kato}\affiliation{\YAMAGATA} 
\author{S.~Kobayashi}\affiliation{\SAGA}
\author{T.~K.~Komatsubara}\affiliation{\KEK} 
\author{A.~S.~Kurilin}\affiliation{\DUBNA} 
\author{E.~Kuzmin}\affiliation{\DUBNA}
\author{A.~Lednev}\affiliation{\IHEP}\affiliation{\CHICAGO} 
\author{H.~S.~Lee}\affiliation{\PUSAN} 
\author{S.~Y.~Lee}\affiliation{\PUSAN} 
\author{G.~Y.~Lim}\affiliation{\KEK}
\author{J.~Ma}\affiliation{\CHICAGO}
\author{T.~Matsumura}\affiliation{\NDA}
\author{A.~Moisseenko}\affiliation{\DUBNA}
\author{H.~Morii}\affiliation{\KYOTO} 
\author{T.~Morimoto}\affiliation{\KEK}
\author{T.~Nakano}\affiliation{\RCNP} 
\author{H.~Nanjo}\affiliation{\KYOTO}
\author{J.~Nix}\affiliation{\CHICAGO}
\author{T.~Nomura}\affiliation{\KYOTO}
\author{M.~Nomachi}\affiliation{\OSAKA}
\author{H.~Okuno}\affiliation{\KEK}
\author{K.~Omata}\affiliation{\KEK}
\author{G.~N.~Perdue}\affiliation{\CHICAGO} 
\author{S.~Podolsky}\altaffiliation{Present address: \GOMEL}\affiliation{\DUBNA} 
\author{K.~Sakashita}\altaffiliation{Present address: \KEK}\affiliation{\OSAKA} 
\author{T.~Sasaki}\affiliation{\YAMAGATA} 
\author{N.~Sasao}\affiliation{\KYOTO}
\author{H.~Sato}\affiliation{\YAMAGATA}
\author{T.~Sato}\affiliation{\KEK}
\author{M.~Sekimoto}\affiliation{\KEK}
\author{T.~Shinkawa}\affiliation{\NDA}
\author{Y.~Sugaya}\affiliation{\OSAKA}
\author{A.~Sugiyama}\affiliation{\SAGA}
\author{T.~Sumida}\affiliation{\KYOTO}
\author{S.~Suzuki}\affiliation{\SAGA}
\author{Y.~Tajima}\affiliation{\YAMAGATA}
\author{S.~Takita}\affiliation{\YAMAGATA} 
\author{Z.~Tsamalaidze}\affiliation{\DUBNA}
\author{T.~Tsukamoto}\altaffiliation{Deceased.}\affiliation{\SAGA} 
\author{Y.~C.~Tung}\affiliation{\TAIWAN}
\author{Y.~Wah}\affiliation{\CHICAGO}
\author{H.~Watanabe}\altaffiliation{Present address: \KEK}\affiliation{\CHICAGO}
\author{M.~L.~Wu}\affiliation{\TAIWAN}
\author{M.~Yamaga}\altaffiliation{Present address: \RIKEN}\affiliation{\KEK}
\author{T.~Yamanaka}\affiliation{\OSAKA}
\author{H.~Y.~Yoshida}\affiliation{\YAMAGATA}
\author{Y.~Yoshimura}\affiliation{\KEK}
\collaboration{E391a Collaboration}\noaffiliation

\preprint{KEK-Preprint/2007-66}

\begin{abstract}
 We performed a search for the $\kpnn$ decay at the KEK 12-GeV 
 proton synchrotron.  No candidate events were observed.
 An upper limit on the branching ratio for the decay
 was set to be $6.7 \times 10^{-8}$ at the 90\% confidence level. 
\end{abstract}

\maketitle

 The decay $\kpnn$ occurs via loop diagrams
that change the quark flavor from strange to down~\cite{Laur}.
 It violates CP symmetry directly, and the amplitude is proportional 
to the imaginary part 
 of the Cabbibo-Kobayashi-Maskawa matrix elements in the standard model (SM).
 Since the theoretical uncertainty in the branching ratio is small and controlled, 
it provides a good testing ground of the SM and beyond~\cite{BSM}.
 The branching ratio $Br(\kpnn)$ is predicted to be 
 $(2.49\pm 0.39)\times 10^{-11}$~\cite{MesciaSmith}.
 The current experimental limit is
 $2.1\times 10^{-7}$ at the 90\% confidence level
 by our previous search~\cite{e391run1}.

KEK E391a is the first experiment dedicated to the $\kpnn$ decay.
Neutral kaons were produced by 12~GeV protons
incident on a 0.8-cm-diameter and 6-cm-long platinum target.
The proton intensity was typically $2 \times 10^{12}$
per spill coming every 4 sec.
The neutral beam~\cite{beamline}, 
with a solid angle of 12.6~$\mu$str,
 was defined by a series of six sets of collimators and a pair of sweeping magnets 
aligned at a production angle of 4~degrees. 
A 7-cm-thick lead block and a 30-cm-thick beryllium block were placed
between the first and second collimators
to reduce beam photons and neutrons.
 The $K^0_L$ momentum 
peaked around 2~GeV/c at the entrance of the detector,
11~m downstream from the target.

\begin{figure}[b]
   \includegraphics[angle=-90, width=8.6cm]{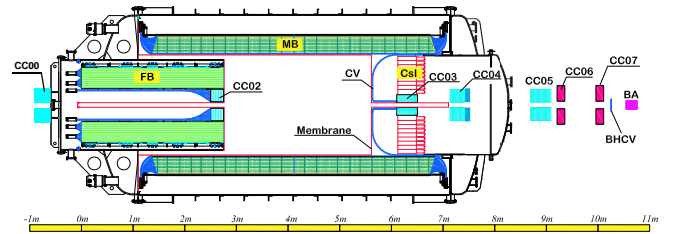}
   \caption{\label{fig:detector}%
	(color online) 
	Schematic cross-sectional view of the E391a detector.
	``0m'' in the scale corresponds to the entrance of the detector.}
\end{figure}

Figure~\ref{fig:detector} shows the cross-sectional view
of the E391a detector. 
 $K^0_L$'s entered from the left side, and
 the detector components were cylindrically assembled
 along the beam axis.
 Most of them were installed inside the vacuum tank 
 to minimize interactions of the particles before detection. 
The electromagnetic calorimeter, labeled ``CsI'',
 measured the energy and position of the two photons from $\pi^0$.
It consisted
 of 496 blocks of \(7 \times 7 \times 30~\mbox{cm}^3\)
 undoped CsI crystal and 80 specially shaped CsI blocks 
 used in the peripheral region, 
 covering a 190~cm-$\phi$ circular area.
To allow beam particles to pass through, 
the calorimeter had a $12 \times 12$~cm$^2$ hole 
at the center.
The main barrel (MB) and 
 front barrel (FB) counters consisted of 
 alternating layers of lead and scintillator sheets with 
total thicknesses of 13.5~$X_0$ and 17.5~$X_0$, respectively,
and surrounded the decay region. 
 To identify charged particles entering the calorimeter, 
scintillation counters (CV) hermetically covered the front of the calorimeter. 
It consisted of a plastic scintillator hodoscope
 that were placed 50~cm upstream of the calorimeter
 and four 6-mm-thick scintillator plates 
 that were located parallel to the beam axis
 between the hodoscope and the calorimeter. 
 Multiple collar-shaped photon counters
 (CC00, CC02--07) were placed 
 along the beam axis
to detect particles escaping 
 in the beam direction.
CC02 was a shashlik type lead-scintillator sandwich counter, and
was located at the upstream end of the $K^0_L$ decay region.
CC03 filled the volume between the beam hole 
and the innermost layers of the CsI blocks in the calorimeter. 
 The vacuum region was separated by a thin multi-layer film (``membrane'') 
 into the beam and detector regions.
This kept the decay region at  $1 \times 10^{-5}$~Pa despite some outgassing from the detector.
Further descriptions of the E391a detector are given in \cite{e391run1,detector}.

 The E391a experiment started taking data in February 2004.
 In the first period, 
 whose partial analysis was reported in \cite{e391run1}, 
 the membrane drooped into the neutral beam near the calorimeter and 
caused many neutron-induced backgrounds.
After fixing this problem,
we resumed the physics run in 2005.
 In this analysis, 
 we used the data in the second period from February to April 2005. 
Data were taken with a hardware trigger requiring two or 
  more shower clusters in the calorimeter with $\ge 60$~MeV.  We also required 
no activity in the CV and in some other photon counters.

 Analysis procedures were 
 basically the same as described in \cite{e391run1}.
First, we identified clusters in the calorimeter, 
 each of which should have
 transverse shower shape consistent with a single photon.
 The clusters were required to have more than 150 and 250~MeV for
 the lower and higher energy photons from $\pi^0$, respectively.
 To improve the accuracy of energy measurement,
 clusters within the $36 \times 36$~cm$^2$ square around the beam and
 outside the radius of 88~cm
 were not used as the photon candidates. 
Second, we selected events with exactly two photons in the calorimeter and 
 without any in-time hits in the other counters. 
 In order to achieve 
 high efficiency of particle detection~\cite{vetoineff},
 energy thresholds for the counters were
 set at around 1~MeV; \mbox{e.g.} 1.0~MeV for FB, MB, and CC02, and 0.3~MeV for CV.
 An additional photon counter (BA) was placed in the beam at the downstream end. 
 It consisted of a series of alternating layers of lead, quartz, and scintillator plates.
 Photons to the BA were identified by 
 the \v{C}erenkov light in the quartz layers 
  and by the energy deposition of more than 20~MeV in the scintillator layers.
Third, assuming that two photons came from a $\pi^0$ decay on the beam axis,
 we calculated the decay vertex position along the beam axis (Z) 
 and the transverse momentum of $\pi^0$ ($P_T$).
Fourth, we imposed kinematic requirements as follows. 
 To remove $\kgg$ decays, 
 we calculated the opening angle between two photon directions projected on the calorimeter plane,
 and required it to be $\le 135$~degrees.
 The shower shape was required to be consistent with 
 a photon entering the calorimeter 
 with the direction from the decay vertex 
 to the hit position on the surface. 
 The reconstructed $\pi^0$ should have the energy less than 2~GeV,
 and should be kinematically consistent with 
 a $\kpnn$ decay within the proper $K^0_L$ momentum range.
Finally, we defined the region for the candidate events (signal box) in the $P_T$ vs Z plot
 as $0.12<P_T<0.24$~GeV/c and $340<Z<500$~cm.
 In this analysis, we masked the signal box
 so that all the selection criteria (cuts) were determined
 without examining the candidate events. 

There were two types of background events. 
One was the events from $K^0_L$ decays and
the other was the events due to 
     the neutrons in the halo of the neutral beam (``halo neutrons'').

\begin{figure}[htb]
   \includegraphics[width=7.8cm]{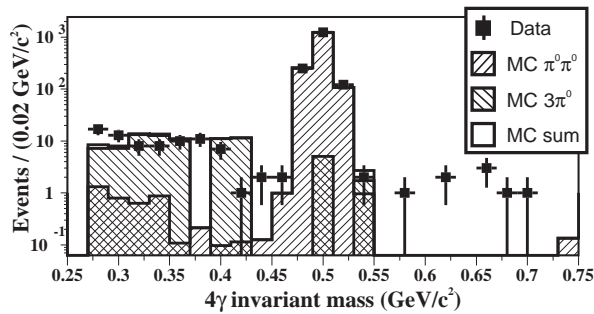}
   \caption{\label{fig:4g}%
	Reconstructed invariant mass distribution of the events with four photons in the calorimeter. 
   The points show the data, and the histograms indicates 
   the contribution of $\kpp$ and $\kppp$ decays 
   (and their sum),
   expected from the simulation,
	normalized with the number of events in the $\kpp$ peak.
}
\end{figure}

 The main background source from $K^0_L$ decays was the $\kpp$ mode,
 whose branching ratio is $8.7 \times 10^{-4}$.
 There are four photons in the final state, and if two of them escape detection,
$\kpp$ can fake a signal event.
The number of background events 
 was estimated by Monte Carlo simulation.
  We generated $\kpp$ decays with 11 times larger statistics than the data.
After imposing all the cuts, 
the background level was estimated to be 0.11 events.
It turned out that the CsI calorimeter and the main barrel were
the most responsible components for detecting extra photons and rejecting those backgrounds.
 To verify the detection inefficiency of photon counters
 in the simulation, 
 we analyzed the events with four photons reconstructed in the calorimeter. 
As shown in Fig.~\ref{fig:4g},
 in addition to the $\kpp$ events at the $K^0_L$ mass, there was a tail in the lower mass region 
due to contamination from $\kppp$,
 whose two out of six photons escaped detection. 
The number of events in the tail was reproduced by our simulation%
~\cite{highmass}.
 For charged decay modes ($\kcppp$, $\kpln$),  we studied 
the rejection power of the kinematic cuts to these backgrounds.
Multiplying the expected inefficiency of charged particle counters to their rejection,
we estimated their contribution to be negligible. 

\begin{figure}[htb]
   \includegraphics[width=7.6cm]{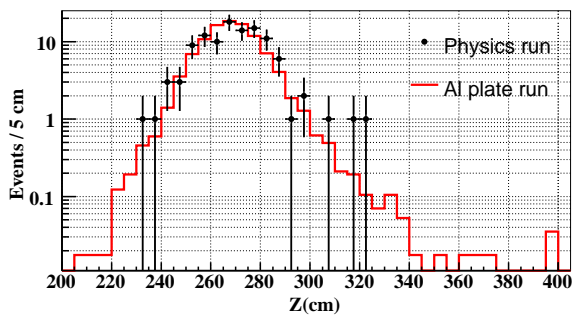}
	\caption{\label{fig:cc02}%
	(color online) 
	Reconstructed Z vertex distribution of $\pi^0$'s produced within CC02.
	The points show the data in the physics run in the upstream region ($Z<340$~cm)
	and the histogram indicates the distribution from the Al plate run.}
\end{figure}

Halo neutrons induced a substantial portion of backgrounds,
    although the halo was suppressed by 5 orders of magnitude from the beam core.
    The background was categorized into three types and they were estimated separately.
The first type was due to $\pi^0$'s produced  
in the interaction of halo neutrons with the upstream 
CC02 collar counter (``CC02 BG'').   Ideally, their Z position should be 
 reconstructed properly at CC02, {\it i.e.}, outside the signal box.  However, they can 
 enter the signal region when the energy of either photon was mismeasured
 due to shower leakage or photo-nuclear interactions in the calorimeter.  To reproduce the tail in the 
vertex distribution, 
 we used data obtained in a dedicated run for the study (``Al plate run''), 
 in which a 0.5-cm-thick aluminum plate was inserted to the beam
at 6.5~cm downstream of the rear end of CC02~\cite{corehalo}.
 After imposing analysis cuts
 and selecting events with two photons 
 in the calorimeter 
 whose invariant mass was consistent with $\pi^0$, 
 we obtained the distribution of reconstructed Z vertex of $\pi^0$'s 
produced at the Al plate.
It was then convoluted with the Z distribution of $\pi^0$'s production points
within CC02~\cite{ZinCC02}
so as to match the peak position 
 with that observed in the physics run, 
as shown in Fig.~\ref{fig:cc02}.
The distribution was normalized to the number of events in $Z<300$~cm.
We estimated the number of CC02 BG events
inside the signal box to be 0.16.

 The second type of neutron-induced background was 
due to neutron interactions with the CV (``CV BG''). 
 This background should also be reconstructed properly at the Z position of the CV, 
{\it i.e.}, outside the signal box.
 However, events can shift upstream when either cluster was overlapped by 
other associated particles
and thereby mismeasured, 
  or when one of the clusters (or both)
  was in fact not due to a photon from $\pi^0$. 
 In order to evaluate the background level inside the signal box,
 we performed a bifurcation study with data~\cite{e949bifurcation, Jons}.
 In the simulation studied beforehand, the cuts against extra particles 
 and the shower shape cut turned out to be efficient in the background reduction;
 these cuts were chosen as two 
uncorrelated cut sets in the bifurcation study.  
The rejection power of one cut set was evaluated with inverting another 
 cut set, and vice versa.  Multiplying the obtained rejection factors, 
the number of CV BG events  
inside the signal box was estimated to be 0.08. 

\begin{figure}[htb]
   \includegraphics[width=7.6cm]{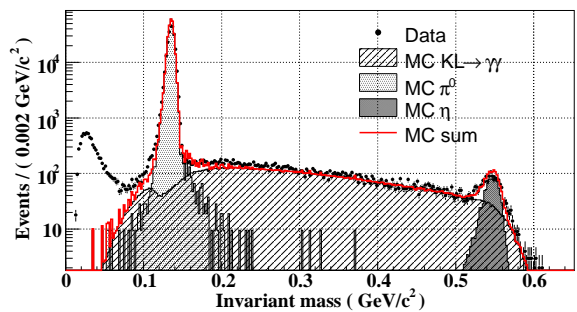}
   \caption{\label{fig:eta}%
	(color online) 
	Reconstructed invariant mass distribution of the two photon events 
        in the Al plate run, explained in the text.
	Points with error bars show the data.
	Histograms indicate the contributions from $\pi^0$ and $\eta$ 
	produced in the Al plate, $\kgg$ decays, and their sum, respectively, from the simulation. 
	Events in the low mass region were considered to be due to neutron interactions 
	accompanying neither $\pi^0$'s nor $\eta$'s,  which were not recorded in the simulation.}
\end{figure}

The third type of neutron-induced background was due to $\eta$'s produced by
the halo through interactions with the CV (``CV-$\eta$ BG'').
 Since the Z vertex position was calculated by assuming the $\pi^0$ mass, 
$\eta$'s were reconstructed about four times 
farther away from the calorimeter,
 and they can fall into the signal box.  To simulate the $\eta$ production, we used a 
GEANT4-based simulation with the Binary Cascade hadron interaction model~\cite{geant4}. 
 Figure~\ref{fig:eta} demonstrates the simulation,
 which reproduced the invariant mass distribution 
 (from $\pi^0$ mass to $\eta$ mass)
 of the events with two photons 
 in the calorimeter
 from the Al plate run,
 normalized by the number of protons on the target.
We then simulated $\eta$ production at the CV and 
estimated the number of CV-$\eta$ BG  events 
inside the signal box to be 0.06.

Table~\ref{table:BG} summarizes the estimated numbers of background events
inside the signal box.
\begin{table}[tb]
   \caption{\label{table:BG}%
	Estimated numbers of background events (BG) inside the signal box.}
   \begin{ruledtabular}
   \begin{tabular}{lD{!}{\,\pm\,}{6,6}}
   Background source & \multicolumn{1}{c}{Estimated number of BG}\\
   \hline
   $\kpp$   &  0.11 ! 0.09 \\
   \hline
   CC02    &  0.16 ! 0.05 \\
   CV       &  0.08 ! 0.04 \\
   CV-$\eta$  &  0.06 ! 0.02 \\
   \hline
   total      &  0.41 ! 0.11 \\
   \end{tabular}
   \end{ruledtabular}
\end{table}
We also examined the numbers of events 
observed in several regions around the signal box, 
and they were statistically consistent with the estimates.

\begin{figure}[b]
   \includegraphics[width=6cm]{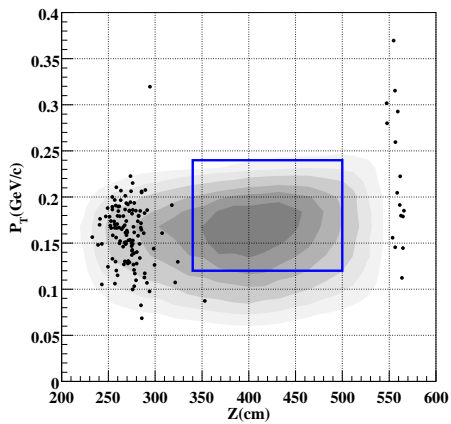}
   \caption{\label{fig:finalplot}%
	(color online) 
	Scatter plot of $P_T$ vs reconstructed Z position after imposing all the cuts.
    The points show the data and 
    the contour represents the simulated distribution of the signal. 
    The rectangle indicates the signal region.}
\end{figure}

After determining all the selection criteria and estimating background levels,
 we examined the events in the signal box and found no candidates,
as shown in Fig.~\ref{fig:finalplot}.

The number of collected $K^0_L$ decays was estimated using the $\kpp$ decay, 
based on 1495 reconstructed events, and
was cross-checked by measuring $\kppp$ and $\kgg$ decays~\cite{PDG}.
The 5\% discrepancy observed between these modes was accounted for as 
an additional systematic uncertainty.
The single event sensitivity for the $\kpnn$ branching ratio is given by
\begin{eqnarray}
	\nonumber
    S.E.S.(\kpnn) = \frac{1}{\mbox{Acceptance}\cdot N(K^0_L \mbox{decays})} \; ,
\end{eqnarray}
where the acceptance includes the geometrical acceptance, the analysis efficiency, and
the acceptance loss due to accidental hits.
Using the total acceptance of 0.67\% and the number of $K^0_L$ decays of $5.1\times 10^{9}$,
the single event sensitivity was $(2.9\pm 0.3) \times 10^{-8}$,
where the error includes both statistical and systematic uncertainties.

Since we observed no events inside the signal box,
we set an upper limit for the $\kpnn$ branching ratio,
\begin{eqnarray}
	\nonumber
   Br(\kpnn) < 6.7 \times 10^{-8} \; (90\% C.L.) \; ,
\end{eqnarray}
based on the Poisson statistics. 
In deriving the limit,
the uncertainty of the single event sensitivity was not taken into consideration.
The result improves the previous limit~\cite{e391run1} by a factor 3,
and the background level by an order of magnitude.

\begin{acknowledgements}
We are grateful to the continuous support by KEK and 
the successful beam operation by the crew of the KEK 12-GeV proton synchrotron. 
This work has been partly supported by a Grant-in-Aid from the MEXT and JSPS in Japan, 
a grant from NSC in Taiwan, a grant from KRF in Korea,  and the U.S. Department of Energy.
\end{acknowledgements}

\bibliographystyle{plain}

\begin{thebibliography}{99}
\bibitem{highmass} 
Events in the higher mass region come from $\kppp$ with mis-combination(s) of photons.
The discrepancy is due to the small sample size of the simulation.
\bibitem{corehalo}
 The neutron spectra in the core and halo regions of the beam were different.
 However, it was found that, 
 when two photons from $\pi^0$ were detected in the calorimeter, 
 the energy and transverse momentum distributions 
 of the $\pi^0$'s produced at the CC02 by the core neutrons
 were similar to those of the $\pi^0$'s by the halo neutrons.
\bibitem{ZinCC02}
We confirmed by simulation studies  
the Z distribution of $\pi^0$'s produced within CC02 was expressed 
as an exponential of the distance from the rear end of the CC02.
\end{thebibliography}

\end{document}